\magnification=1095

\input eplain
\input epsf
\input amssym.def
\input amssym
 
\ifx\tenpoint\undefined\let\loadedfrommacro=Y%
\ifx\loadedfrommacro Y\else
         \message{10point.TeX must be loaded from a macro package.}
         \message{Input terminated.}
          \fi
 
\font\tencsc=cmcsc10
 
\newfam\scfam
 
\def\tenpoint{\def\rm{\fam0\tenrm}
    \textfont0=\tenrm  \scriptfont0=\sevenrm  \scriptscriptfont0=\fiverm
    \textfont1=\teni   \scriptfont1=\seveni   \scriptscriptfont1=\fivei
    \textfont2=\tensy  \scriptfont2=\sevensy  \scriptscriptfont2=\fivesy
    \textfont3=\tenex  \scriptfont3=\tenex    \scriptscriptfont3=\tenex
    \textfont\itfam=\tenit   \def\it{\fam\itfam\tenit}%
    \textfont\slfam=\tensl   \def\sl{\fam\slfam\tensl}%
    \textfont\ttfam=\tentt   \def\tt{\fam\ttfam\tentt}%
    \textfont\bffam=\tenbf   \scriptfont\bffam=\sevenbf
    \scriptscriptfont\bffam=\fivebf  \def\bf{\fam\bffam\tenbf}%
    \textfont\scfam=\tencsc  \def\sc{\fam\scfam\tencsc}%
    \normalbaselineskip=12pt
    \setbox\strutbox=\hbox{\vrule height8.5pt depth 3.5pt width0pt}%
    \normalbaselines\rm}

         \let\loadedfrommacro=N\fi
\font\ninerm=cmr9            \font\sixrm=cmr6
\font\ninei=cmmi9            \font\sixi=cmmi6
\font\ninesy=cmsy9           \font\sixsy=cmsy6
\font\ninebf=cmbx9           \font\sixbf=cmbx6
\font\ninesl=cmsl9           \font\ninett=cmtt9      \font\nineit=cmti9
\font\ninecsc=cmcsc10
\font\ninebfit=cmbxti10 at 9pt
\ifx\ninepoint\undefined
   \def\ninepoint{\def\rm{\fam0\ninerm}
       \textfont0=\ninerm  \scriptfont0=\sixrm  \scriptscriptfont0=\fiverm
       \textfont1=\ninei   \scriptfont1=\sixi   \scriptscriptfont1=\fivei
       \textfont2=\ninesy  \scriptfont2=\sixsy  \scriptscriptfont2=\fivesy
       \textfont3=\tenex   \scriptfont3=\tenex  \scriptscriptfont3=\tenex
       \def\bfit{\ninebfit}%
       \textfont\itfam=\nineit   \def\it{\fam\itfam\nineit}%
       \textfont\slfam=\ninesl   \def\sl{\fam\slfam\ninesl}%
       \textfont\ttfam=\ninett   \def\tt{\fam\ttfam\ninett}%
       \textfont\bffam=\ninebf   \scriptfont\bffam=\sixbf
        \scriptscriptfont\bffam=\fivebf   \def\bf{\fam\bffam\ninebf}%
       \textfont\scfam=\ninecsc  \def\sc{\fam\scfam\ninecsc}%
       \normalbaselineskip=11pt%
       \setbox\strutbox=\hbox{\vrule height8pt depth3pt width0pt}%
       \normalbaselines\rm}%
   \fi

\input colordvi

%


\newcount\boguscount
\boguscount=12
\catcode`@ = \letter
\let\@eplainoldatcode=\boguscount
\catcode`@=\letter

\newcount\footnotenum \footnotenum=0

\def\footnote#1{\let\@sf\empty 
  \ifhmode\edef\@sf{\spacefactor\the\spacefactor}\/\fi
  $^{#1}$\@sf\vfootnote{#1}}
\def\vfootnote#1{\insert\footins\bgroup
  \interlinepenalty\interfootnotelinepenalty
  \splittopskip\ht\strutbox 
  \splitmaxdepth\dp\strutbox \floatingpenalty\@MM
  \leftskip\z@skip \rightskip\z@skip \spaceskip\z@skip \xspaceskip\z@skip
  \ninepoint
  \setbox2=\hbox{#1.\enspace}%
  \hangindent=\wd2 \hangafter=0 
  \noindent \llap{\box2}\footstrut\futurelet\next\fo@t}

\catcode`@ = \@eplainoldatcode

\long\def\ft#1#2{\global\advance\footnotenum by 1%
\definexref{#1}{\the\footnotenum}{footnote}%
  \footnote{\the\footnotenum}{#2}}

\font\medbf=cmbx12

\def\verythinback{\mskip-1.5mu}
\def\edot#1{\verythinback\cdot\verythinback\e{#1}}

\def\e#1{10^{#1}}

\footline={\hfil}
\headline={\hfil\ifnum\pageno>1\bf\folio\fi}

\advance \baselineskip by 1pt

\font\medbf=cmbx12
\font\sc=cmcsc10

\listrightindent=15pt
\interitemskipamount=1.5pt

\def\myref#1{%
\edef\temp{\getproperty{\xrlabel{#1}}{class}}%
\textRed\ref{#1}\textBlack%
}

\def\myrefs#1{%
\edef\temp{\getproperty{\xrlabel{#1}}{class}}%
\textRed\refs{#1}\textBlack%
}

\def\myrefn#1{%
\edef\temp{\getproperty{\xrlabel{#1}}{class}}%
\textRed\refn{#1}\textBlack%
}

{\catcode`\#=11%
\global\def\numbersign{#}}

\def\eqlabel{\eqdef}
\def\neqno{\eqdef{}}

\def\ifig#1#2#3{%
\global\advance\figureNum by 1%
\definexref{#2}{\the\figureNum}{figure}%
\midinsert
\begingroup
\leftskip=\parindent
\parindent=0pt
\halign{##\hfil\cr
\leftline{\vbox{\epsfbox{#1}}}\cr \noalign{\smallskip}
\vbox{\ninepoint \hsize=5truein%
\noindent {\ninebfit \noindent
Figure~\the\figureNum.}~{\it #3}}\cr}
\par
\endgroup
\endinsert}

\newcount\sectionNum \sectionNum=0
\newcount\subsectionNum \subsectionNum=0

\def\newsectionNum{\global\advance\sectionNum by 1 \global\subsectionNum=0
\relax \thesectionNum\relax}
\def\newsubsectionNum{\global\advance\subsectionNum by 1 \relax
\thesubsectionNum\relax}

\def\thesectionNum{\the\sectionNum}
\def\thesubsectionNum{\the\sectionNum.\the\subsectionNum}

\def\section#1 \par{\bigbreak\noindent
{\bf \llap{\newsectionNum~}%
#1%
}
\writenumberedtocentry{section}{#1}{\the\sectionNum}%
\nobreak\smallskip\noindent\ignorespaces}

\def\sectionl#1#2 \par{\section{#1} \par
\definexref{#2}{\thesectionNum}{section}}

\def\subsection#1 \par{%
\bigbreak\noindent
{\it \llap{\newsubsectionNum~}%
#1%
}%
\writenumberedtocentry{subsection}{#1}{\the\subsectionNum}%
\nobreak\vskip1.5pt\noindent\ignorespaces}

\def\subsectionl#1#2 \par{\subsection{#1} \par
\definexref{#2}%
{\thesubsectionNum}{section}}

\newcount\figureNum \figureNum=0
\newbox\figbox \newdimen\xsize \newskip\ysize \newbox\realfigbox
\newbox\figboxtemp \newbox\tempbox

\def\grad{\triangledown}

\def\bv{{\bf v}}
\def\bdot{{\bf\cdot}}
\def\p{\partial}
\def\vdg#1#2{v_{#1}{\p v_{#2}\over\p{#1}}}
\def\onelap#1#2{{\p^2 v_{#2}\over\p{#1}^2}}
\def\laplong#1{\onelap x{#1} + \onelap y{#1} + \onelap z{#1}}
\def\piece#1{\left(\vdg x{#1} + \vdg y{#1} + \vdg z{#1}\right) + 
{\p v_{#1}\over\p t} &= -{1\over\rho}{\p p\over\p{#1}} +
\nu\left(\laplong {#1}\right)}

\def\litre{\,\ell}
\def\x{\times}
\def\N{\u{N}}
\def\calories{\u{cal}}
\def\uof#1{\left[#1\right]}
\def\cd{c_{\rm d}}
\def\RE{{\sl Re}}
\def\cgsnu{\nuunits}
\def\and{\quad\hbox{and}\quad}
\def\M{{\rm M}}
\def\L{{\rm L}}
\def\T{{\rm T}}

\def\mytitle#1{\vbox{\parindent=0pt
{\medbf #1}\medskip {\sc Sanjoy Mahajan}\par
Cavendish Laboratory \par
Astrophysics\par
Cambridge CB3 0HE \par
England\par
\mailurl{sanjoy@mrao.cam.ac.uk}\par}\bigskip}


\def\rdelimit{$\rangle$}
\def\ldelimit{$\langle$}

\def\rawurlaux#1#2%
{\ldefs
\leavevmode%
\tdefs
\textRed #2%
\textBlack
\egroup}

\def\urlaux#1{\rawurlaux{#1}{\ldelimit #1\rdelimit}}
\def\mailurlaux#1{\rawurlaux{mailto:#1}{#1}}

\begingroup
\catcode`_=13 
\catcode`\&=13
\catcode`\$=13
\catcode`-=13
\catcode`/=13
\gdef\fixu{\def_{\leavevmode \kern.06em \vbox{\hrule width.3em}}}
\catcode`.=13
\gdef\tdefs{\fixu
\def&{\char38}%
\def.{\discretionary{\char46}{}{\char46}}%
\def-{\discretionary{\char45}{}{\char45}}%
\def/{\discretionary{\char47}{}{\char47}}%
\def~{\char126}}

\gdef\ldefs{\chardef\_=`\_ \let_=\_%
\let&=\&%
\let$=\$%
\chardef\~=`\~%
\let~=\~%
\chardef\.=`.%
\let.=\.%
\chardef\-=`-%
\let-=\-%
\chardef\/=`/%
\let/=\/%
}

\gdef\rawurl{\bgroup\catcode`~=13%
\catcode`_=13%
\catcode`\&=13%
\catcode`\$=13%
\rawurlaux}

\gdef\url{\bgroup\catcode`~=13%
\catcode`_=13%
\catcode`\&=13%
\catcode`\$=13%
\catcode`.=13%
\catcode`=13%
\catcode`/=13%
\urlaux}

\gdef\mailurl{\bgroup\catcode`~=13%
\catcode`_=13%
\catcode`\&=13%
\catcode`\$=13%
\catcode`.=13%
\catcode`=13%
\catcode`/=13%
\mailurlaux}

\endgroup


\def\u#1{{\rm\,#1}}


\def\gm{\u{g}}
\def\g{\gm}
\def\kg{\u{kg}}


\def\m{\u{m}}
\def\mm{\u{mm}}

\def\cm{\u{cm}}
\def\km{\u{km}}

\def\A{\u{\AA}}


\def\seconds{\u{s}}

\def\yr{\u{yr}}





\def\eV{\u{eV}}

\def\cm{\u{cm}}

\def\J{\u{J}}

\def\K{\u{K}}

\def\W{\u{W}}


\def\litre{\,\ell}




\def\mps{\m\seconds^{-1}}
\def\cmps{\cm\seconds^{-1}}
\def\mph{\u{mph}}


\def\mpss{\m\seconds^{-2}}


\def\nuunits{\cm^2\seconds^{-1}}






\mytitle{Estimating gas mileage: An example of order-of-magnitude physics}

\begingroup \advance\baselineskip by -1pt
\noindent {\it Based on a talk, `Lying and estimating for general
education', at the 121st AAPT National Meeting, Guelph, Ontario, 31 July 2000.}
\bigskip

\hrule\smallskip
\noindent {\bf Abstract.}  I discuss how to estimate the gas mileage
of a car.  This discussion, which covers air resistance and Reynolds
numbers, describes one way to introduce dimensional
analysis and order-of-magnitude physics into introductory physics (if
only the syllabus would allow it).  It is part teacher's guide and
part textbook chapter -- I hope not the worst parts of each.
\smallskip\hrule
\endgroup

\bigskip\bigskip\bigskip
\leftline{\sc Contents}\bigskip

\begingroup

\newdimen\secspace
\newdimen\subsecspace

\newcount\secnum
\newcount\subsecnum

\let\subsecfont=\it
\let\secfont=\rm

\def\numgap{\hskip0.6em\relax}
\def\endgap{}

\setbox0=\hbox{\secfont 0\numgap}
\secspace=\wd0

\setbox0=\hbox{\subsecfont 0.0\numgap}
\subsecspace=\wd0

\parindent=0pt

\def\tocsubsectionentry#1#2#3{%
\line{\hskip\secspace
\textRed\rm\the\secnum.#2\textBlack
\numgap#1\dotfill \rm #3\endgap}}


\readtocfile

\endgroup
\vfill
{\ninepoint
\noindent Copyright \copyright\ 2001--2005 by Sanjoy Mahajan.
Licensed under the Open Software License version 3.0. This document is
free/open-source software.  See the file COPYING in the source code.}
\eject

\section{The problem}

Can we predict the gas mileage for a car (in miles per gallon)?  We
can begin the discussion by asking students why a car requires
gasoline.  Where does the energy go?  Eventually students say: some
sort of resistance.  What kind?  Air resistance.  Here is a chance to
teach a principle of science: {\it Test your ideas}.  Have confidence
in your ideas, but not too much; the arms-control negotiator says
`trust, but verify'.  We test our model -- that air resistance
consumes most of the power -- by calculating whether air resistance
accounts for the gasoline consumed.

\def\Egal{E_{\rm gallon}}

How large is air resistance?  Before students can answer `how large',
they must think about how to measure air resistance.  Is it a force, a
pressure, an energy?  Gasoline provides energy, so let's compute the
energy consumed by air resistance, and equate it to the energy
provided by one gallon of gasoline.  Energy is force times distance:
$E_{\rm drag}=Fd$, where $F$ is the air-resistance force and $d$ is
distance traveled.  If $\Egal$ is the energy provided by one gallon of
gasoline, and $\Egal\sim E_{\rm drag}$, then $d=\Egal/F$ is the
distance a car can travel on that gallon.  The problem breaks into two
computations: the air-resistance force and the energy available from 1
gallon of gasoline.  This breakdown is an example of {\it
divide-and-conquer\/} reasoning, a frequent technique in
order-of-magnitude physics and in everyday thinking.

\section{Air resistance}

How can we compute the air-resistance force?  We can scare students by
writing down the Navier--Stokes equations from fluid mechanics, as a
vector equation with gradients and dot products:
$$(\bv\bdot\grad)\bv + {\p\bv\over\p t} = -{1\over\rho}\grad p +
\nu\grad^2\bv.\neqno$$
If the plethora of symbols confuses students, consider it a job well
done.  Now we can increase the tension, when we tell them
that these equations are vector
shorthand for three coupled nonlinear partial-differential equations:
$$\eqalign{\piece x,\cr
\piece y,\cr
\piece z.\cr}\eqlabel{Navier--Stokes-full}$$
To find the force, we solve these equations
for the pressure, $p$.  We'll solve this problem after studying
partial-differential equations for three years.
Students with any imagination
by now tremble a bit, and are receptive to a simpler method.
When they hear that we have not listed the complete set of
equations -- the set \eqref{Navier--Stokes-full}
leaves out the continuity equation --
students are distressed.  Estimation plus dimensional analysis
is a simple and quick method for finding the drag force.

\subsection{Choosing relevant quantities}

These approximate methods, although mathematically simple, require
physical imagination.  To stimulate the imagination, we being by
deciding which features of the problem determine the air resistance.
Air, like any fluid, resists the motion of an object moving through
it.  This description suggests two categories of relevant features:
characteristics of the car and of the air.

The car's speed, $v$, determines drag.  Gales can knock over trees;
gentle breezes cannot.  This argument about moving air might cause
students to wonder: Weren't we talking about still air and a moving
car?  We were, but the two descriptions -- moving air with stationary
car, or stationary air with moving car -- are equivalent.  Therefore,
we can reason about a reference frame in which a stationary car is
buffeted by a wind (of speed $v$), and transfer that reasoning to the
frame where the car moves through still air.

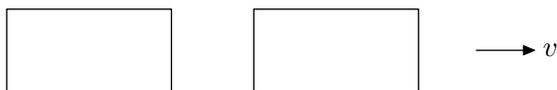
\ifig{fig.1}{fig:tailgate}{Two cars, one tailgating the other
(view from the side).}

The car's size also matters: Large cars feel more drag than small cars
do.  How should we measure size?  Length should not affect air
resistance, as the following thought experiment suggests.  Imagine two
cars, one tailgating the other (\myref{fig:tailgate}). The rear car
feels almost no drag; some cyclists try a related
dangerous activity: riding behind a truck to reduce air resistance (as
in the movie {\it Breaking Away\/} (1979)).
In the limit of zero tailgating distance (\myref{fig:zero-tailgate}),
the two cars merge into one long car.  The long car
has the same drag as one short
car.  

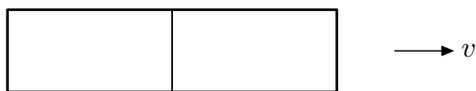
\ifig{fig.2}{fig:zero-tailgate}{No  distance between the
cars (view from the side).  The two short cars meld into one long car
(heavy outline)
that feels the same air resistance as one short car feels.}

This last statement is surprising, but you can perform a demonstration
to convince yourself and your students.  Hold a book in one hand and a
piece of paper in the other hand, at say chest height.  Ask which
object will hit the ground first.  Most predict that the book hits
first.  Don't drop anything, for that would only reward rash
responses!  Instead place the paper under the book (choose the paper
so that it's slightly smaller than the book) and then drop the
combined object.  They hit at the same time.  The audience will
protest that you cheated, because `the book is forcing the paper
down'.  Agree with the criticism: Offer to put the paper on top of the
book and drop the book and paper.  However, ask for predictions first:
What will happen?  The two objects fall as one.  Many dubious
explanations will be offered, including that the book `sucks the paper
downward'.  But the simplest explanation is also the correct one: The
top object (the paper) feels no air resistance, so it falls like a
stone.  The bottom object (the book) feels air resistance, but being
heavy the drag hardly affects it on the short journey to the floor (it
too falls like a stone).  Similarly, in \myref{fig:zero-tailgate}, the
second car experiences no air resistance, so the double car feels the
same drag as one short car does.  Therefore, car length should not
affect air resistance.

Physicists experiment in their minds all the time.  Some theorists
are, like me, limited to thought experiments, where equipment is cheap
and clumsiness no handicap.  Experimentalists also use thought
experiments; how else could they design a real experiment?  Skill in
designing and using such experiments is one of the most valuable
lessons that physics can teach.  It develops the student's
imagination.  I therefore interleave thought experiments throughout
this discussion.

\ifig{fig.3}{fig:side-by-side}{Two cars traveling side by side
(view from above).}

To decide how width affects air resistance, consider a related thought
experiment: two cars traveling side by side (\myref{fig:side-by-side}).
Each car feels the same resistance as one car.  In the limit that the
two cars are adjacent, the double-width car feels twice the resistance
of one single-width car.  So resistance should be proportional to
width.  A thought experiment with one car traveling above the other
suggests that resistance should be proportional also to height.  So a
reasonable measure of size is height times width, or frontal area $A$.
The analysis of the the relevance of area shows students two examples
of thought experiments.  Three examples are a charm: Students
understand an idea after seeing three examples that use it.  Read on
to see the third example.

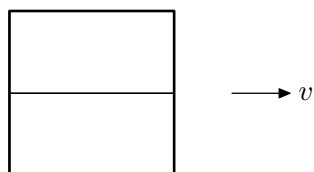
\ifig{fig.4}{fig:side-by-side-limit}{Limit of two cars traveling
side by side (view from above).  The two cars meld into one wide car
(heavy outline)
that experiences twice the air resistance that one thin car feels.}

The density of the fluid also determines drag.  If students don't
realize that the density of air matters, ask why it is tiring to run in
a swimming pool: because water is thick and air is thin.  When we
discuss density of the fluid, students naturally wonder whether the
density of the car affects air resistance.  To answer this question,
we can use another thought experiment -- the third!  Imagine a car
with its windows sealed, traveling at $60\mph$.  Stop the car, 
invite four large friends
into the car -- preferably friends raised on steak,
potatoes, and growth hormone -- and speed up to $60\mph$.  The density
of the car increases, but does the air know about the contents of the
car?  No.  To the air, the car is a black box: Its contents are
invisible.  The air knows only the car's speed and the shape and
texture of the its surface.  So the density of the car should not affect
air resistance.

Our thought experiments tell us that the drag force, $F$, depends on
$\rho$, $v$, and $A$.  It could also depend on viscosity, a reasonable
proposal since viscosity is the only mechanism of energy loss in the
problem, so it is the only source of drag.  If the viscosity is
exactly zero, then the drag is also zero.  However, as long as the
viscosity is not zero, the drag depends only slightly on the
viscosity.  The explanation is tricky, and the simplest route around
this obstacle is to say, `Trust me for now that the viscosity does not
matter.  We'll derive our result, then do an experiment at the end to
check whether it is reasonable, and thereby check whether I deserve to
be trusted on this point.'  For readers of a less trusting
disposition, jump ahead to \myrefs{sec:stokes-law} and
\myrefn{sec:reynolds-number}, which discuss the relative importance of
viscous and inertial drags, and justify the neglect of viscosity.

We often tell students that dimensions are part of a physical
quantity, rather than an extra, like salt, to add according to taste.
But students do not understand why we exhort them on this point.  Here
we can show them: finding the drag merely by requiring that $F$ have
dimensions of force.

\subsection{Dimensions of each quantity}

What are the dimensions of each variable?  Students know that force can be
measured in Newtons, but they often do not realize what dimensions a Newton
contains.  So we remind them that any valid equation for force, such
as $F=ma$, determines
the dimensions: $$\uof{F}=\M\L\T^{-2}.\neqno$$  The dimensions of
speed and area cause no trouble:
$$\eqalign{\uof{v}&=\L\T^{-1}\cr
\uof{A}&=\L^2.\cr}\neqno$$
Nor should the dimensions of density:
$$\uof{\rho}=\M\L^{-3}.\neqno$$
But some students think that density is volume per mass.  They
memorized the phrase `mass per volume' badly, and did not learn the important
idea: that an ice cube and an iceberg have the same density, that
density is {\it intensive}.  When we discuss the dimensions of density, we
can distinguish intensive quantities, such as density and
temperature, from extensive quantities, such as mass and heat.

\subsectionl{Looking for the right combination}{sec:right-combination}

How can we combine these variables into a quantity with the dimensions
of force?  When many teach what they call dimensional analysis, they
show students how to set up and solve linear equations in order to
find the right combination: count powers of mass, length, and time in
each variable -- so each variable becomes a three-dimensional vector
in the space of dimensions -- and ask what linear combination of
$\rho$, $v$, and $A$ vectors makes a force vector.  This problem is
equivalent to solving a system of linear equations.  I like reasoning
using the space of dimensions; I should use it to argue intuitively
for the Buckingham Pi theorem (quoted without proof in
\cite[Chapter~3]{mahajan:98} and used many times in the rest of the
document; see Buckingham's paper \cite{Buckingham:1914} for the
original statement and proof).  But solving the linear equations is
pointless.  It is a brute-force method that teaches the student little
except how to solve linear equations.  If a problem is so complicated
that we must solve linear equations to find the right combination,
then we have too many variables; dimensional analysis will not save
us.  We need first to simplify the list of variables by using
additional physical arguments.

Instead of solving linear equations,
we can teach a quick and elegant method.  Force contains one
power of mass; the only other variable that contains mass is $\rho$,
which also contains one power of mass.  So $F$ must be proportional to
$\rho$.  Now the problem simplifies: How to combine $v$ and $A$ into
$F/\rho$, which has dimensions of $\rm L^4T^{-2}$.  Apply the same
trick to time: $F/\rho$ contains time as $\T^{-2}$, and only the speed has
time in it.  The speed contains time as $\T^{-1}$, so $F/\rho \propto
v^2$.  The problem is now even simpler: What do we do to $A$ to make a
quantity with the dimensions of $F/\rho v^2$?  The dimensions of $A$
and $F/\rho v^2$ are the same, so $F/\rho v^2\!A$ is dimensionless.
This method of constraints is subtle (it
substitutes thought for mindless calculation), but reasoning with
constraints is valuable for analyzing complicated problems and is worth
teaching.  In finding the drag formula, students use the
method twice with slight variations.  Repetition teaches, but
repetition with variation teaches more.  (Poly\`a \cite{Polya:62} 
points out that
Mozart, in his piano concertos, did not merely repeat the theme;
rather, Mozart restated it with variations.)

Since $F/\rho v^2\!A$ is dimensionless, it must be a constant.
Voil\`a: Drag force is proportional to $\rho v^2\!A$, a result that we have
found without solving any differential equations.  Earlier we promised
that 


Could $F$ be
$7000\rho v^2\!A$ or $\rho v^2\!A /1000$?  Sure; our method does not
tell us the constant.  To find the constant, we would have to solve
Navier--Stokes equations \eqref{Navier--Stokes-full}.  

This property is general.  When you solve a differential equation, you
learn only a dimensionless constant; the rest of the solution -- the
functional form -- is determined by physical constraints, the same
constraints that determine the form of the differential equation.  The
simplest method tells you the most important information; Murphy's law
is not often violated, but when it is, we should be grateful!
Differential equations are difficult; physical arguments we can teach.

Let's analyze free fall, the first problem that students solve with
differential equations.  How long does a rock take to fall
from a height of $10\m$ (roughly three storeys)?  The time depends on
the strength of gravity, $g$, and the height, $h$.  How can $g$ and
$h$ combine into a quantity with dimensions of time?  There is only
one way: $t\sim \sqrt{h/g}$.  We can find that expression using the
method of constraints. The input variables $h$ and $g$ each contain
one power of length, and the fall time contains no length, so $t$ must be a
function of the ratio $h/g$:
$$t = f(h/g).\neqno$$  
To decide on the functional form, look at the
powers of time: $h/g$ contains $\T^2$, so 
$$t=\sqrt{h/g},$$
except for a dimensionless constant.  The fall
time from three storeys is roughly $$t\sim
\left(10\m\over10\mpss\right)^{1/2}=1\sec.$$ The differential equation
for the position of the object is
$${d^2 x\over dt^2}=g,\neqno$$ where $x$ is the distance traveled
since release and $t$ is the time since release.  The solution,
$x(t)=gt^2/2$, tells us that the object falls a distance $h$ when
$t=\sqrt{2h/g}$.  The order-of-magnitude analysis left out a
dimensionless factor of $\sqrt2$.  In an order-of-magnitude analysis,
we {\it hope\/} that the missing constant is close to unity, and often
it is.  It is
worth hoping: Solving a differential equation is much
harder than fiddling with dimensions and performing thought
experiments.

We now test our conclusion that $\rho v^2\!A$ a
reasonable expression for drag force.  Drag should increase as density
increases, as speed increases, or as area increases.  Our expression
has these properties.  This test suggests an alternative method that
we could have used to determine the drag force -- an alternative worth
using if students find the constraint method too tricky.  Drag force
should increase with speed, density, and area.  So let's try the
formula: $F\sim \rho v A$.  The dimensions of $\rho v A$ are
$\M\T^{-1}$.  The dimensions of force are $\M\L\T^{-2}$, so our
expressions lacks a factor of $\L\T^{-1}$.  One more power of $v$
fixes this problem, and we find that $F\sim \rho v^2\!A$.

\subsectionl{Stokes' law}{sec:stokes-law}

What if a student looks in her textbook and finds Stokes' law for a
sphere: $$F=6\pi\rho\nu v r,\neqno$$ where $\nu$ is kinematic
viscosity of the fluid and $r$ is the radius of the sphere.  Why
didn't our argument discover Stokes' law?  This question is excellent.
If a student does not raise the question, we can raise it ourselves.
A simple answer is that throwing out viscosity makes it impossible to
discover Stokes' law.  But let's pretend that we didn't throw out
viscosity.  In discussing Stokes' law, we get an excuse to discuss
viscosity and to compare the relative sizes of the inertial ($\rho
v^2\!A$) and Stokes' drag forces.  Their ratio is the simplest
comparison:
$${\hbox{inertial drag force}\over\hbox{Stokes' drag force}}
\sim {\rho v^2\!A\over \rho\nu v r}\sim {vr\over\nu},\eqlabel{RE-force-ratio}$$
where we have estimated the area $A$ as $r^2$.  This ratio is
dimensionless, and is therefore a valuable quantity.  It
is the {\it Reynolds number}, commonly denoted $\RE$, and is 
a measure of
the flow speed (or, equivalently, of the object's speed).  Speed?  We
divided forces; where did speed enter?  In the expressions for the
drag forces.  The Reynolds number turns out to be proportional to
$v$.   Alone $v$ cannot measure speed, because 
$v$ is not dimensionless; its value depends on the system
of units.  I walk at $3\mph$.  To make the speed seem slow, I
can quote it as
$$\eqalign{v_{\rm walk}&= 1.5\edot{-3}\km\sec^{-1}\cr
&=1.5\edot{-9}\u{parsecs}\yr^{-1}.\cr}\neqno$$
To make the speed seem fast, I can quote it as 
$$\eqalign{v_{\rm walk}&= 5\edot{4}\km\yr^{-1}\cr
&=5\edot{21}\A\u{century}^{-1}.\cr}\neqno$$

This example illustrates an important principle: {\bf No quantity with
dimensions is big or small intrinsically.}  Is $5\kg$ a large mass?
For a bacterium, yes; for an elephant, no.  A quantity with dimensions
must be compared to another, {\it relevant} quantity with the same
dimensions; dividing the two quantities results in a dimensionless
number, whose value is independent of the system of units.  In
searching for a relevant comparison, students explore a problem and
connect what they discover to their other knowledge.  If students had
this habit, they would pause before writing down whatever number
appears on their calculator display.  An inclined plane with a height
of $\e{-7}\m$ or a charge of $\e7\u{C}$ would make students suspect a
mistake.

A simple explanation of the Reynolds number is the ratio of inertial
and Stokes' drag expressions, as shown in
\eqref{RE-force-ratio}.  This explanation is slightly misleading.
At high Reynolds number, the Stokes' drag expression does not apply;
at low Reynolds number, the inertial drag expression does not apply.
There's no regime where both expressions apply; taking their ratio is
physically slightly misleading.  But it is a reasonable way to produce
a dimensionless number.

As an alternative explanation, the Reynolds number is the ratio
of the object's speed and $v_{\rm diffuse}=\nu/r$, the speed at which
momentum diffuses.  Kinematic viscosity, $\nu$, is the diffusivity of
momentum; momentum therefore diffuses across an object of size $r$ in
time $t\sim r^2/\nu$ (as a dimensional argument suggests).  From the
length $r$ and the time $t$, we can form a speed:
$$v_{\rm diffuse}={r\over t}\sim {r\over r^2/\nu}={\nu\over
r},\neqno$$ which it is natural to call the diffusion speed.  

\subsectionl{Reynolds number}{sec:reynolds-number}

Students can estimate the Reynolds number for various flows, and we
can discuss the consequences (oily flow for $\RE\ll1$, turbulent flow
for $\RE\gg1$), and show the beautiful
pictures from {\it An Album of Fluid Motion\/} \cite{van-Dyke:82}
or {\it A Gallery of Fluid Motion\/} \cite{Samimy:2003}.

For example, walking across a room,
$$v\sim 200\cmps,\quad \nu\sim0.2\cgsnu,\and r\sim 100\cm,\neqno$$ so
$$\RE\sim {200\cmps\x100\cm\over0.2\cgsnu}\sim \e5.\neqno$$ Or,
running in a swimming pool:
$$v\sim 100\cmps,\quad \nu\sim\e{-2}\cgsnu,\and r\sim 100\cm,\neqno$$ so
$$\RE\sim {100\cmps\x100\cm\over\e{-2}\cgsnu}\sim \e6.\neqno$$
I have
quoted quantities in cgs units rather than in the more common SI (mks)
units, so that students see the arbitrariness of unit systems and do
not become wedded to a single system.

The only tricky part in the preceding estimate is determining $r$
(are you a sphere?).  But we need only an approximate Reynolds number, so an
approximate measure of our size is accurate enough for this estimate.
This Reynolds
number is much greater than unity -- a convenient dividing line between
fast and slow flows -- so the flow is fast.  Experiments show that for
$\RE$ greater than roughly 1000, flow is turbulent.  Because air is
invisible, we do not appreciate the turbulence that we generate
merely by walking, but physics increases the power of our imagination.
The Reynolds number in this example is so large that we expect most
everyday flows to be turbulent as well.

Another example: a paramecium swimming in pond water.  Students can
estimate the speed by putting a drop of pond water under the
microscope and noting how long it takes the little beast to cross the
field of view.  I shall make a rough estimate here, based on hazy
memories of school biology.  At 1000-fold magnification, a paramecium
looks $1\cm$ long, the field of view looks $15\cm$ wide, and the
paramecium swims across it in perhaps $15\sec$.  I had originally
written $30\sec$, but I am hardly confident of either value, so I
might as well use the numerically convenient value of $15\sec$.
The ingredients of the Reynolds number are $$r\sim\e{-3}\cm,\quad v\sim
\e{-3}\cmps,\and \nu\sim\e{-2}\cgsnu,\neqno$$
so the Reynolds number is 
$$\RE\sim{\e{-3}\cmps\x\e{-3}\cm\over\e{-2}\cgsnu}\sim\e{-4}.\neqno$$
The flow is excruciatingly slow and viscous; to the paramecium, water
is a thick, viscous liquid, the way cold honey or corn syrup is to us.
Purcell's article on `Life at low Reynolds number' \cite{Purcell:77}, a
beautiful discussion of this point, is one that we and our students
can enjoy.

For everyday flows, inertial drag is the important drag, which
explains why we won't worry about Stokes' drag for gas mileage (it
turns out that the Stokes'-drag formula is valid only for $\RE\ll1$,
and the inertial-drag formula only for $\RE\gg1$).

\subsection{Checking the expression for inertial drag}

Now we can return to the inertial drag force.  We have already checked
the expression theoretically, when we verified in
\myref{sec:right-combination} that the form was reasonable.  We can also
check it experimentally, by {\it putting in numbers}.  We get another
chance to reinforce the moral: Doubt, question, check, never trust
yourself completely.  Now that we are about to do arithmetic, I tell
students that `calculators rot their brain'.  I forbid my students
from using them; they would be able to calculate to one digit without
a calculator, except that calculator use has atrophied their numerical
sense.  So students need to practice -- they need to put in numbers --
to recover their feel for numbers.

In what situation can we test the formula for the drag force?
Eventually we test it when we estimate the gas mileage, but the
gas-mileage example is not the ideal test: We want to use the formula
to test also whether air resistance is the main contribution to gas
mileage.  So we ought to test the formula in another example -- to
gather independent evidence.  Ideally, this new example would use
students' knowledge of their everyday world.  Students learn little if
we show them how the drag formula constrains, for example, the design
of supersonic transports.  

Instead we might analyze why running in a swimming pool is so
exhausting, and how fast people can run in a swimming pool.  The speed
is limited by the power that a person can generate; this power goes to
fighting drag.  How much power can a person generate?  It depends on
the person, but let's ask about a typical person.  The power is
roughly a few hundred watts -- as a student may know if at a science
museum she has tried to light a bulb using a bicycle.  So $P_{\rm
avail}\sim300\W$.  Always ask, and get students to ask: {\bf How
reasonable is that number?}  One way to judge it is to compare it to
another, similar power: the horsepower, roughly $750\W$.  So a person,
with, say, one-fifth the mass of a horse and presumably one-fifth the
muscle mass too, can put out almost one-half the power?  Maybe the
$300\W$ is an overestimate, but on the other hand, humans have lots of
muscle in their legs, whereas horses have -- for their greater weight
-- relatively spindly legs.  So maybe a hard-cycling human can
generate more power per mass than a horse can, and the $300\W$ is
roughly right.  Either way, it's not far off so let's use the value.

The power consumed by drag is the drag force times the person's speed
or $\rho v^3\!A$.  The estimated speed is $$v\sim \left(P_{\rm
avail}\over
\rho A\right)^{1/3}.\neqno$$
Students now get another chance to put in numbers.  The density of
water is easy: $\e3\kg\m^{-3}$.  My frontal area -- divide-and-conquer
reasoning once again -- is $2\m\x0.5\m$ or $1\m^2$.  To estimate an
area, split the problem in two: into estimating length and estimating
height. Arons, in {\it Teaching Introductory Physics}
\cite[p.~12]{Arons:97}, discusses how students `know' the area of a square or
of a circle, but not of an irregular figure, for which no formula is
available; the notion that area is length times width, even when the
length and width are not precisely defined, does not occur to
students.  An order-of-magnitude area estimate, such as for a person's
frontal area, teaches this idea.

We now put the pieces together to find $v$:
$$v\sim
\left(300\W \over
\e3\kg\m^{-3}\x1\m^2\right)^{1/3}.\eqlabel{pool-speed}$$
As soon
as we write down this expression, students reach for their calculators
-- an opening for us to wax eloquent on the evils of
calculators, and to show how to do the calculation by hand.  We write
300 as $0.3\edot3$; then the powers of ten
cancel, leaving only $0.3^{1/3}\mps$.  So $v\sim1\mps$,
which is $2\mph$.  A useful approximation: $1\mps\sim2\mph$.  Is this
speed reasonable?  Yes -- when I run in water, I cannot keep up with
someone strolling alongside on the edge of the pool (a typical walking
speed is $3\mph$).  The agreement with everyday experience
increases our confidence in the drag formula.  We can also point out
that, even if we estimated the $P_{\rm avail}$ inaccurately (and we
probably did), the error in the speed is small because of the blessed
one-third power in the speed expression \eqref{pool-speed}.

\subsection{Drag force for a car}

Emboldened, we use the drag formula for the original question, gas
mileage.  What is the frontal area for a car?  A car is not as tall as
a person, so the height is $1.5\m$.  When I go car camping and sleep
in the back seat, I fit but do not consider it luxury accommodation; so
the car's width is maybe $1.5\m$.  The area is therefore
$1.5\m\x1.5\m\sim2\m^2$.  To estimate the speed, pick a typical
highway speed: $60\mph$, or $30\mps$.

There are many ways to estimate the density of air.  One
method is to remember that $22\litre$ is one mole at
standard conditions (sea-level pressure and room temperature).  Air is
mostly dinitrogen ($\hbox{N}_2$), with a molecular weight of 28. So $22\litre$ has a
mass of $28\g$.  The density is roughly $1\g\litre^{-1}$ or
$1\kg\m^{-3}$.  A more involved method derives the $22\litre$ magic
number from the ideal gas law.  For one mole, $PV=RT$, where $P$ is
pressure, $V$ is the volume of one mole, $R$ is the gas constant, and
$T$ is the temperature.  We can look up the gas constant, and we
know the temperature.  Atmospheric pressure is easy to remember from
(American) weather reports: `Barometer is 30 inches and falling'.  One inch is
$25\mm$, so atmospheric pressure is equivalent to a column roughly
$750\mm$ high.  But $750\mm$ of what?  Of mercury.  Mercury is 13
times denser than water, so atmospheric pressure is equivalent to a
column of water roughly $13\x750\mm$ high, or $h\sim10\m$.  The resulting
pressure is, from hydrostatics, 
$$P=\rho g h \sim \e3\kg\m^{-3}\x10\mpss\x10\m=\e5\N\m^{-2}.\neqno$$
The volume occupied by one mole of atmosphere is
$$V = {RT\over P}\sim {8\J\K^{-1}\x 300\K\over \e5\N\m^{-2}}\sim
24\litre.\neqno$$ This calculation is another one that students can
do mentally.  The method is simple: Do the important parts first.  So
we first count the powers of 10.  Rewrite `8' as $0.8\edot1$; then
there are three powers of 10 in the numerator, and five in the
denominator, which combine into $\e{-2}$.  The remaining factors are
small and easy to handle mentally: $0.8\x3$, or $2.4$.  So the volume
is $2.4\edot{-2}\m^3$ or $24\litre$.  This method of determining the
molar volume, which starts with the ideal gas law, shows students how
much they can estimate without looking up many quantities.  Such
estimation develops number sense and connects otherwise
disparate bits of physics.

We now have computed the numbers that we need to estimate the drag force:
$$F\sim1\kg\m^{-3}\x(30\mps)^2\x2\m^2\sim 2\edot3\N.\neqno$$ This
mental calculation is simple using the identity $30\x30=1000$.  Other
useful order-of-magnitude rules of arithmetic include
$$\eqalign{2\x2\x2&=10,\cr
4\x4&=20,\cr
\pi&=3.\cr}\neqno$$

\section{Energy of gasoline}

How much energy does a car get from 1 gallon of gasoline?  What is
this absurd unit, the gallon?  It is 4 quarts; each quart is roughly
one liter, so for our purposes, 1 gallon is $4\litre$.  But $4\litre$
of what?  Gasoline is like fat in the energy that it stores.  The
nutrition information on the back of a soup can tells us that fat
gives $10$ calories per gram or $4\edot4\J\g^{-1}$.  [Let the students
use that number (converting incorrectly to Joules at
$4\calories\J^{-1}$) and complete the calculation.  When they compute
a horribly low mileage, ask why.  Eventually students realize that
nutritional calories are {\it kilocalories}.  They can then redo the
calculation using the proper conversion.]  A favorite question: How
reasonable is this value?  To judge it, we should get a second
opinion, for example from chemistry.  Most chemical reactions release
a few $\hbox{eV}$ per molecule.  For a long-chain hydrocarbon like
gasoline, a molecular unit, say $\hbox{CH}_2$, might be a better basis
for that calculation.  The molar mass of $\hbox{CH}_2$ is $14\g$, so
the energy density would be:
$${3\eV\x6\edot{23}\over14\g}\x{1.6\edot{-19}\J\over1\eV}\sim
2\edot4\J\g^{-1}.$$
Given the uncertainty in the pieces of this calculation, it agrees
reasonably well with the soup-label estimate of $4\edot4\J\g^{-1}$.
What is the mass of $4\litre$ of gasoline?  In the
order-of-magnitude world, every liquid is water, so $4\litre$ has a
mass of $4000\g$.  Its energy content is
$4\edot4\J\g^{-1}$, so the energy provided by 1 gallon is
$$E_{\rm gallon}\sim 4000\g\x4\edot4\J\g^{-1}\sim
2\edot8\J,\neqno$$ where the last step follows from the `identity'
$4\x4=20$.

\section{Mileage}

The energy that the car requires is the drag force times the distance
traveled, $d$.  Thus $E_{\rm avail}=Fd$.  The distance traveled is
$$d\sim {E\over F}\sim {2\edot8\J\over2\edot3\N}\sim \e5\m,\neqno$$ or
$100\km$.  Our prediction -- a 60 miles-per-gallon car -- is
reasonable.

We got a bit lucky.  The drag is roughly one-fourth of what we
estimated; the formula leaves out a factor of $0.5\cd$, where $\cd$ is
the drag coefficient (typically 0.5 for most cars -- I once saw an ad
for a sports car that quoted 0.33).  The efficiency of the engine is
not 1.0, but more like 0.25, which is also the efficiency of human
metabolism.  The two errors canceled, and we got an unreasonably
accurate value.  But that cancellation shows another advantage of 
order-of-magnitude 
methods: If you split the problem into enough parts, the errors in the
different parts may cancel!

Our mileage estimate is reasonable, so we have answered our original
question: Air resistance does cause a significant amount of the total
resistance, at least at highway speeds.  This analysis suggests a
follow-up question: How much extra oil would the United States require
if everyone drove $80\mph$ instead of $60\mph$ on the highway?

\section{Acknowledgments}

Many thanks to David Hogg for detailed, insightful comments.

\vfill\eject
\section{References}

\vskip -\lastskip
\bibliography{mileage}
\bibliographystyle{plain}

\end